# Electronic structure and transport properties of sol-gel-derived high-entropy Ba(Zr$_{0.2}$Sn$_{0.2}$Ti$_{0.2}$Hf$_{0.2}$Nb$_{0.2}$)O$_3$ thin films

Yuqi Liang[1], Bingcheng Luo[1a)], Huijuan Dong[1], Danyang Wang[2]

[1]School of Physical Science and Technology, Northwestern Polytechnical University, Xi'an, 710072, China

[2]School of Materials Science and Engineering, The University of New South Wales, Sydney, NSW 2052, Australia

a) Electronic mail: luobingcheng@nwpu.edu.cn

## Abstract

High-entropy perovskite thin films, as the prototypical representative of the high-entropy oxides with novel electrical and magnetic features, have recently attracted great attention. Here, we reported the electronic structure and charge transport properties of sol-gel-derived high-entropy Ba(Zr$_{0.2}$Sn$_{0.2}$Ti$_{0.2}$Hf$_{0.2}$Nb$_{0.2}$)O$_3$ thin films annealed at various temperatures. By means of X-ray photoelectron spectroscopy and absorption spectrum, it is found that the conduction-band-minimum shifts downward and the valence-band-maximum shifts upward with the increase of annealing temperature, leading to the narrowed band gap. Electrical resistance measurements confirmed a semiconductor-like behavior for all the thin films. Two charge transport mechanisms, *i.e.*, the thermally-activated transport mechanism at high temperatures and the activation-less transport mechanism at low temperatures, are identified by a self-consistent analysis method. These findings provide a critical insight into the electronic band structure and charge transport behavior of Ba(Zr$_{0.2}$Sn$_{0.2}$Ti$_{0.2}$Hf$_{0.2}$Nb$_{0.2}$)O$_3$, validating it as a compelling high-entropy oxide material for future electronic/energy-related technologies.

**Introduction**

Following the high-entropy alloys (HEA) proposed in 2004 by Yeh *et al.* [1], the high-entropy concept was extended to other material systems ranging from carbides to oxides [2-5]. Among them, high-entropy oxides (HEO) are an emerging class of materials that display a broad range of novel functionalities. Since the first report by Rost *et al.* on entropy-driven structural stabilization of single-phase $(Mg_{0.2}Co_{0.2}Ni_{0.2}Cu_{0.2}Zn_{0.2})O$ [3], low thermal conductivity [6], colossal dielectric properties [7], increased electrical storage capacities and lithium ionic conductivity [8] have been observed in HEOs. These tantalizing functionalities in HEOs have greatly stimulated the research enthusiasm on their fundamental properties and potential applications.

As a member of the HEO family, perovskite oxides are particularly attractive given their overwhelming advantages including highly tolerant ions and a wealth of functionalities from superconductivity to multiferroism. In the first attempt of synthesizing high-entropy perovskite oxide ceramics in 2018, Jiang *et al.* demonstrated that the single-phase structure can be formed at a specific sintering temperature (~1300 $^0C$) [9]. Despite a limited number of recent studies, high-entropy perovskite oxides exhibited many interesting properties including ultralow thermal conductivity [6], antiferromagnetic behavior [10-12], and temperature-insensitive permittivity [13], making them promising candidates for potential applications in microelectronics, energy storage and conversion, data storage. Notably, the electronic structure of high-entropy perovskite oxides, which is highly critical for realizing applications involving energy and opto-electronic devices, has not been studied thoroughly. On the other hand, the highly dispersed cations in high-entropy perovskite oxides could incur the lattice distortion and increase the degree of disorder due to the different cation size, and thus strongly affect the transport behavior, which is of importance for energy-related applications [14].

In this work, thin films of a prototypical high-entropy perovskite oxide namely Ba(Zr$_{0.2}$Sn$_{0.2}$Ti$_{0.2}$Hf$_{0.2}$Nb$_{0.2}$)O$_3$ (B5) were synthesized by a sol-gel method. The electronic structure and transport properties of the thin films annealed at different temperatures were investigated. The sol-gel method is of great interest for growth of high-entropy oxide thin films owing to its advantages over vacuum-depositions, such as better stoichiometric control, lower thermal treatment temperature, higher uniformity over a large area and better compatibility with flexible substrates [15].

**Experimental**

High-entropy perovskite oxide thin films Ba(Zr$_{0.2}$Sn$_{0.2}$Ti$_{0.2}$Hf$_{0.2}$Nb$_{0.2}$)O$_3$ (B5) were deposited on MgO (001) single crystal substrates (HeFei Crystal Technical Material Co., Ltd.) by a sol-gel method. The starting materials including high-purity barium acetate [Ba(CH$_3$COO)$_2$], zirconium butoxide [Zr(OC$_4$H$_9$)$_4$], tin chloride dihydrate [SnCl$_2$·2H$_2$O], hafnium chloride [HfCl$_4$], niobium ethoxide [Nb(OC$_2$H$_5$)$_5$], lactic acid [C$_3$H$_6$O$_3$], 2-Methoxyethanol (CH$_3$OCH$_2$CH$_2$OH), acetic acid (CH$_3$COOH), and acetylacetone (CH$_3$COCH$_2$COCH$_3$), were used to prepare the precursor solution. All these chemicals were purchased from Sigma-Aldrich. The detailed process for precursor solution is given in Figure S1 (See Supplementary Materials). The concentration of the precursor solution is 0.1 mol/L. Before deposition, the prepared solution was filtered using a syringe filter with a size of ~200 nm and aged for >48h. The clear and transparent precursor solution exhibiting Tyndall effect was obtained, as shown in Figure S2.

The B5 thin films were spin-coated on MgO (001) substrates. The spin speed and the spin time were fixed at 5500 rpm and 30 s, respectively. After spin-coating each layer, drying and pyrolysis were performed at 130 and 400 °C for 5 min, respectively. The coating-heating treatments were repeated three times, and then the resultant films were annealed at various temperatures (600, 700 and 800℃) in air atmosphere for 120 min. The thin films are denoted by S600, S700 and S800, respectively.

The phase structures of B5 thin films were investigated by X-ray diffraction (XRD) (PANalytical Empyrean) with Cu $K_α$ radiation in the usual $θ$-$2θ$ geometry. The film thickness and surface elemental mapping of B5 thin films were characterized by a scanning electron microscope (SEM, FEI Verios G4) equipped with an energy dispersive spectroscopy detector. As shown in Figure S3 and S4, the atomic ratio of constituent elements in B5 thin films is consistent with the expected composition, and the thickness of B5 thin films is about 95 nm. The element valence states and valence spectra of B5 thin films were checked using X-ray photoelectron spectroscopy (XPS) (ESCALAB 250) with Al $K_α$ radiation. Absorption spectra were recorded by U3010 spectrophotometer with an integrating sphere in the wavelength range of 250-800 nm using MgO substrate as the reference. The temperature-dependent sheet resistance (*R-T* curves) was measured along the in-plane direction using a Keithley picoammeter/voltage source (model 6487).

**Results and discussion**

Figure 1 (a) shows XRD patterns of B5 thin films. Except the diffraction peak from MgO substrate, the other diffraction peaks match to a perovskite oxide standard pattern (JCPDS No: 31-0174) well, confirming the pure perovskite structure with (00*l*)-preferential orientation [9].

Additionally, at a low annealing temperature, *i.e.*, 600 °C, the diffraction peaks from film were weak, indicating the partial crystallization in S600.

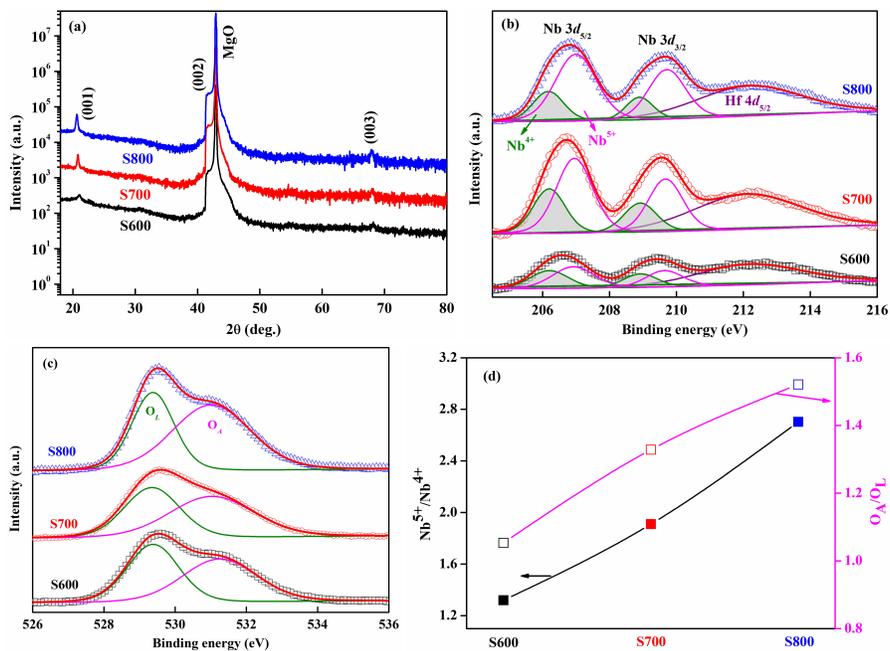

**Figure 1 Structural and chemical characterization of B5 thin films.** (a) XRD patterns, (b) Narrow-scan XPS spectra for Nb 3*d*, (c) Narrow-scan XPS spectra for O 2*p*, (d) Relative concentration ratios of $Nb^{5+}/Nb^{4+}$ and $O_A/O_L$ for samples annealed at different temperatures. The solid lines in (b) and (c) represent the fitting results.

The valence states of constituent elements in B5 thin films were investigated subsequently using X-ray photoelectron spectroscopy (XPS) (ESCALAB 250) with Al $K_\alpha$ radiation. All the expected elements are present, as shown in Figure S5, consistent with the SEM-EDX results (Figure S3). The $Ba^{2+}$, $Zr^{4+}$, $Ti^{4+}$, $Hf^{4+}$, and $Sn^{4+}$ with a fixed valence state can be readily determined regardless of annealing temperatures, and the corresponding XPS analysis are given in Figure S6-S10, respectively. However, valence states of Nb and O elements are quite sensitive to the annealing temperatures. As shown in Figure 1 (b), apart from the peak of Hf $4d_{5/2}$, each peak for Nb 3d, *e.g.*, Nb 3d$_{5/2}$, can be deconvoluted into two components, centered at 206.2±0.1 eV and 207.0±0.1 eV, which can be assigned to Nb$^{4+}$ and Nb$^{5+}$ [16, 17], respectively. Meanwhile, the ratio of the area of Nb$^{5+}$ peak to the area of Nb$^{4+}$ peak increases with the increase of annealing temperature, as shown in Figure 1 (d), implying that the excess free electrons could be provided by Nb$^{5+}$ in the samples annealed at higher temperatures.

Similarly, as shown in Figure 1 (c), the asymmetric O 1s XPS peaks of B5 thin films can also be deconvoluted into two components, centered at 529.4±0.1 eV and 531.1±0.1 eV, respectively. The spectral component appeared at the low binding energy (~529.4 eV) is ascribed to the lattice oxygen (O$_L$), while the other one is associated with the surface-absorbed oxygen (O$_A$) [18]. It is found that the ratio of peak area between O$_A$ and O$_L$ is positively correlated with the annealing temperature, as shown in Figure 1 (d). The evolution of O$_A$ concentration reflects the fact that the excess carriers originating from Nb$^{5+}$ can facilitate surface oxygen absorption to maintain charge neutrality, which will be beneficial for gas sensors and catalysis applications.

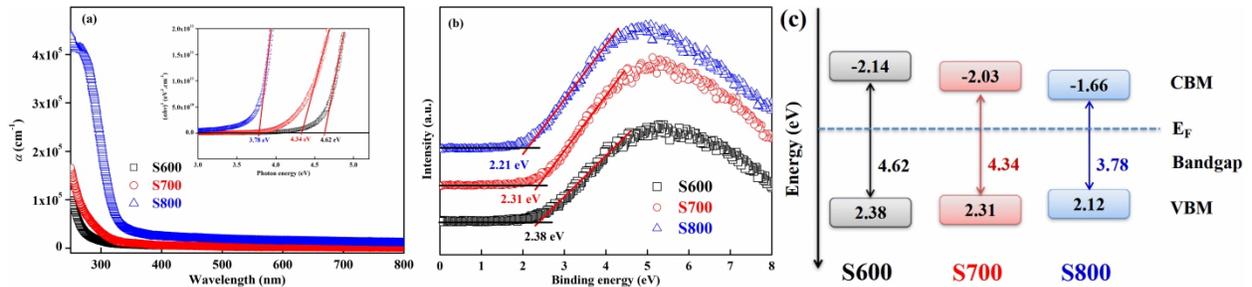

**Figure 2 Electronic structure analysis for B5 thin films.** (a) Optical absorption spectra, (b) Valence-band XPS spectra, (c) Schematic of band structures. Inset in (a) shows the Tauc plots. The solid lines in (a) and (b) represent the fitting results.

We further proceed to examine the electronic band structure of B5 thin films by combination of optical absorption spectra and valence band XPS spectra. Optical absorption spectra ($α$) of B5 thin films exhibit the red-shift of absorption edge with the increase of annealing temperature, as shown in Figure 2 (a), indicating the narrowed band gap. Such narrowed band gap in the sol-gel-derived oxide thin films is commonly attributed to quantum confinement effect [19, 20]. For high-entropy oxides, the existence of multivalent elements, *e.g.*, Nb in the present case, can also cause the narrowed band gap [20]. The optical band gaps ($E_g$) could be estimated by the Tauc plots [21], as shown in inset of Figure 2 (a). The $E_g$ values were determined to be 4.62, 4.34 and 3.78 eV for S600, S700, and S800, respectively. Moreover, the Urbach energy ($E_U$) reflecting the degree of disorder could be calculated through the Urbach relation $α \propto \exp(E/E_U)$ [22], where $E$ is the photon energy, as shown in Figure S11. It is noted that the $E_U$ values (>360 meV) of B5 thin films are larger than those of amorphous perovskite oxide thin films with single cation in B-site (*e.g.*, $E_U$ ~330 meV for amorphous $BaSnO_3$ thin films) [22], which may be the indication of the highly-dispersive-cations induced disorder in high-entropy oxides.

The energy difference between the valence band maximum ($E_{VBM}$) and the Fermi energy ($E_F$, here $E_F$=0) could be determined using the valence band XPS spectra. As shown in Figure 2 (b), from the intersections of two fitting straight lines representing the up edge of the spectrum and the low-energy flat distribution portion, the values of ($E_{VBM}$- $E_F$) were obtained to be 2.38, 2.31 and 2.12 eV for S600, S700, and S800, respectively. Furthermore, the energy difference between the conduction band minimum ($E_{CBM}$) and $E_F$, which is identical to ($E_g$−$E_{VBM}$), was calculated to be 2.14, 2.03 and 1.66 eV for S600, S700 and S800, respectively. Accordingly, the energy band diagrams can be sketched, as schematically shown in Figure 2 (c). Evidently, with the increase of annealing temperature, the CBM moves downwards whereas the VBM shifts upwards, with respect to the Fermi level, leading to the narrowed band gap. Although the key orbitals/elements that contribute to the VBM/CBM is currently unclear for high-entropy perovskite oxides, it is commonly believed that both the orbital interaction and bond-length will influence the locations of the VBM and CBM. The more significant displacement of CBM compared with that of VBM in response to the change in annealing temperature in the present case may be presumably attributed to the presence of multivalent Nb element, resulting in the variation in orbital hybridization and bond-length [23]. Moreover, from the energy band diagram, it is evident that the position of the Fermi level is above the mid-gap, suggesting that the prepared B5 thin films are *n*-type semiconductors. Such a feature is reasonable if one considers the fact that the $Nb^{5+}$ ions act as donors that could provide the excess electrons, as revealed similarly in simple perovskite oxides containing $Nb^{5+}$ ions [24, 25].

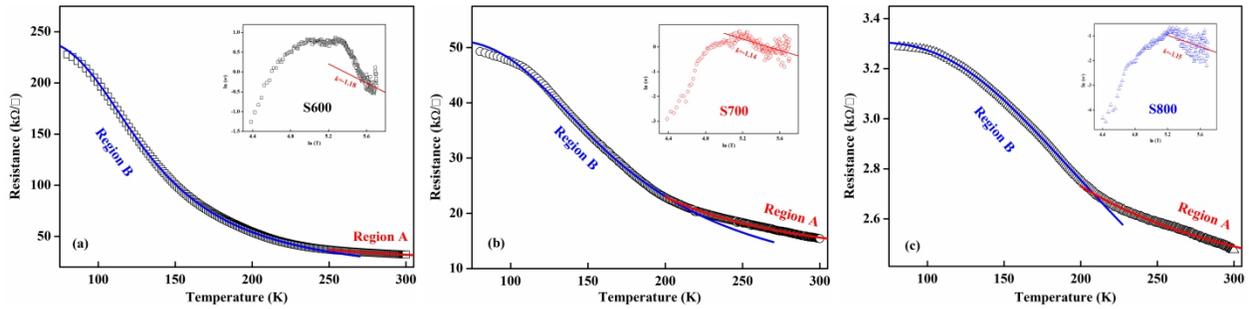

**Figure 3 Transport behaviors of B5 thin films.** Temperature-dependent sheet resistance of (a) S600, (b) S700 and (c) S800. The solid lines represent the fitting results according to the thermally-activated model (red solid lines) and the activation-less model (blue solid lines). Insets in (a), (b) and (c) show the corresponding ln$w$ versus ln$T$ curves, where w = $-d\ln(R)/d\ln(T)$.

Finally, the electronic transport behavior of B5 thin films was determined by measuring the temperature-dependent sheet resistance (*R-T* curves) along the in-plane direction using a Keithley picoammeter/voltage source (model 6487). As shown in Figure 3, the decrease in sheet resistance with temperature in the temperature range of 80 to 300 K is evident, reflecting the semiconducting behavior for all the B5 thin films, which is in good agreement with the above energy band diagrams. The sheet resistance at 300 K is about 32.1, 16.8, and 2.5 kΩ/ for S600, S700, and S800, respectively. The decreased sheet resistance here further confirms the fact that more $Nb^{5+}$ ions are present in the B5 thin films as annealing temperature goes up, echoing the XPS results in Figure 1(b). The obtained sheet resistance of B5 thin films also implies the presence of subgap states involving $Nb^{5+}$ ions that behaves as effective shallow donors. Additionally, the change in the slope of *R-T* curves (Region A and B as marked in Figure 3) in the measuring temperature range indicates that more than one carrier transport mechanisms exist in our high-entropy B5 thin films.

To understand the origin of the electronic transport behaviors, a self-consistent method was used here through calculating the logarithmic derivative $w=-\mathrm{dln}(R)/\mathrm{dln}(T)=-T/R\times \mathrm{d}(R)/\mathrm{d}(T)$ [26]. According to typical electronic transport mechanisms including thermally activated (TA) model, variable-range hopping (VRH) model and adiabatic small polaron (ASP) model in semiconducting perovskite oxide thin films [26, 27], the resistance satisfies a general relation [27]:

$$R=R_0 T^q \exp[(E_a/k_B T)^p] \qquad (1)$$

Where $R_0$ is the pre-exponential term, $k_B$ is the Boltzmann constant, $T$ is the temperature and $E_a$ is the activation energy. $q$ and $p$ are conduction-mechanism-specific constants, *e.g.*, $q=0$ and $p=1$ for the TA conductivity, $q=0$ and $p=0.5$ (0.25) for the Mott (Efros-Shklovskii) VRH conductivity, and $q=p=1$ corresponds to the ASP conductivity [27]. Accordingly, one can expect a negative slope in the ln($w$) vs. ln($T$) curves, *e.g.*, -1 for TA conductivity and -0.5 (-0.25) for Mott (Efros-Shklovskii) VRH conductivity.

The ln($w$) vs. ln($T$) plots are shown in the insets of Figure 3. The negative slopes are evidently noticed at high temperatures, referring to region A, and the positive slopes are seen at low temperatures, referring to region B. Although poor linearity is seen in Region A due to the sensitivity of $w$ values to the scattering in the raw data in a differentiation method, the negative slopes are determined to be ~ -1, suggesting that the TA model dominates the electronic transport behavior. It is also worth mentioning that, for S600, in a certain temperature range in region A the slope is less than -1, which implies that the ASP model may be involved. The ASP mechanism has been found in many oxides containing multi-cations, wherein the electronic carriers move slowly (mobility<1 $cm^2V^{-1}s^{-1}$) to displace the surrounding atoms and thus leading to the formation of polaron [14]. The hypothesis for the ASP model in S600 could be supported by the Hall measurement results, as given in Table S1 in the Supplementary Materials. Moreover, the relation $R=R_0T \exp(E_a/k_BT)$ was used to fit the experimental data in the Region A. As shown in Figure 3 (red solid lines), the fitting curves satisfy the experimental data well and the resulting parameters including $R_0$ and $E_a$ are given in Table S2. The activation energy $E_a$ decreases from 27.3 to 6.6 meV when annealing temperature increases from 600 ºC to 800 ºC. Smaller $E_a$ values suggest a thermally activated conductivity between donor sites in the impurity band rather than through an energy gap from donors to the conduction band [28]. On the other hand, in region B, the slopes of ln($w$) vs. ln($T$) plots are basically positive, which is commonly considered as a sign of metallic conduction. In contrast to typical metallic materials, the sheet resistance of B5 thin films decreases with an increase in temperature following a general temperature-dependent power law, $R \propto T^\beta$ ($0 \leq \beta \leq 1$) [29]. However, no satisfactory fitting results to this power law were obtained (results not shown). With this said, such a unique transport behavior observed in our B5 thin films further stresses the vital role of the highly-dispersive-cation induced disorder in manipulating the electrical properties of high-entropy perovskite oxide thin films. Our B5 thin films possess a high carrier concentration (>$10^{19}$ cm$^{-3}$, Table S1) near the metal-insulator transition, whereas the omnipresent disorder will suppress the metallic feature, collectively resulting in a semiconductor-like *R-T* behavior. Accordingly, it is reasonable to assume that the conduction in region B is governed by a macroscopic percolation system consisting of a metallic channel and a TA-type channel. Thus, the resistance can be described by [30]

$$R^{-1}=A+B\exp(-C/T) \qquad (2)$$

Where *T* is the temperature, *A, B,* and *C* are constants. Indeed, as shown in Figure 3 (blue solid lines), the fitting curves are in good agreement with the experimental data. The parameters *A, B,* and *C* from fitting are given in Table S2. All the three parameters increase when annealing temperature rises, confirming the enhanced metallic conductivity.

**Conclusions**

In summary, the prototypical high-entropy thin films of Ba(Zr$_{0.2}$Sn$_{0.2}$Ti$_{0.2}$Hf$_{0.2}$Nb$_{0.2}$)O$_3$ were prepared by a sol-gel route, and the impact of annealing temperature on their electronic structure and charge transport properties were investigated. With the increase of annealing temperature from 600 ºC to 800 ºC, the conduction-band-minimum shifts downwards whereas the valence-band-maximum shifts upwards, leading to a narrowing of the band gap from 4.62 eV to 3.78 eV. The temperature-dependence of sheet resistance exhibits a semiconductor-like behavior for all the thin films. The enhancement of conductivity with increasing annealing temperature is attributed to the increased concentration of Nb$^{5+}$. Through a self-consistent analysis method, two distinct charge transport mechanisms, namely thermally-activated transport mechanism at high temperatures and activation-less transport mechanism at low temperatures, are identified. This work will serve as a base for future studies on electronic structure and carrier transport of a wide variety of high-entropy oxides, and facilitate the potential applications of high entropy oxides in electronic/energy-related devices.

**Acknowledgments**

This work is partially supported by the Fundamental Research Funds for the Central Universities (No: 310201911cx024) and the National Key Research and Development Program of China (2017YFB0503300).

**Supplementary data**

See the Supplementary Information for the experimental details of film preparation and characterization results of XPS, SEM, optical and electrical measurements.